\documentclass[10pt,journal,compsoc]{IEEEtran}
\IEEEoverridecommandlockouts

\ifCLASSOPTIONcompsoc
  \usepackage[nocompress]{cite}
\else
  \usepackage{cite}
\fi
\ifCLASSINFOpdf
\else
\fi

\usepackage[utf8]{inputenc}
\usepackage{hyperref}
\usepackage[dvipsnames]{xcolor}
\usepackage{cite}
\usepackage{graphicx}
\usepackage{amsmath}
\usepackage{algorithmic}
\usepackage{relsize}
\usepackage{tikz}
\usepackage{pgfplots}
\usepackage{acronym}
\usepackage{tikz}
\usetikzlibrary{shapes,arrows}
\usepackage{pgfplots}
\usepackage{pgfplotstable}
\usepackage{tabularx}
\usepackage{booktabs}
\usepackage{url}
\usepackage{subcaption}

\usepackage{pifont}

\usepackage{xargs}
\usepackage[colorinlistoftodos,prependcaption]{todonotes}
\newcommandx{\unsure}[2][1=]{\todo[linecolor=red,backgroundcolor=red!25,bordercolor=red,#1]{#2}}
\newcommandx{\change}[2][1=]{\todo[linecolor=blue,backgroundcolor=blue!25,bordercolor=blue,#1]{#2}}
\newcommandx{\info}[2][1=]{\todo[linecolor=OliveGreen,backgroundcolor=OliveGreen!25,bordercolor=OliveGreen,#1]{#2}}
\newcommandx{\improvement}[2][1=]{\todo[inline, linecolor=Plum,backgroundcolor=Plum!25,bordercolor=Plum,#1]{#2}}
\newcommandx{\thiswillnotshow}[2][1=]{\todo[disable,#1]{#2}}

\begin{document}
	
\tikzstyle{block} = [draw, rectangle, minimum height=3em, minimum width=6em]
\tikzstyle{edge} = [draw, ->, line width= 0.1em]
\tikzstyle{sum} = [draw, circle, node distance=1.5cm]
\tikzstyle{input} = [coordinate]
\tikzstyle{output} = [coordinate]
\tikzstyle{pinstyle} = [pin edge={to-,thin,black}]

%
\title{Cysteine post-translational modifications: ten years from chemical proteomics to\\ bioinformatics}
%
%
%
%


\author{\IEEEauthorblockN{1\textsuperscript{st} Yanzheng Meng}
	\IEEEauthorblockA{\textit{School of Basic Medicine} \\
		\textit{Qingdao University}\\
		 Qingdao 266021, China \\
		mengyanzheng@genomics.cn}
	
	\and
	\IEEEauthorblockN{1\textsuperscript{st} Lei Li \textsuperscript{$\star$}, \thanks{ Lei Li  is corresponding author.}}
	\IEEEauthorblockA{\textit{School of Basic Medicine} \\
		\textit{Qingdao University}\\
		 Qingdao 266021, China \\
		lileime@hotmail.com}
}
\IEEEtitleabstractindextext{%
\begin{abstract}
As the only thiol-bearing amino acid, cysteine (Cys) residues in proteins have the reactive thiol side chain, which is susceptible to a series of post-translational modifications (PTMs). These PTMs participate in a wide range of biological activities including the alteration of enzymatic reactions, protein-protein interactions and protein stability. Here we summarize the advance of cysteine PTM identification technologies and the features of the various kinds of the PTMs. We also discuss in silico approaches for the prediction of the different types of cysteine modified sites, giving directions for future study.
\end{abstract}

}

\maketitle

\IEEEdisplaynontitleabstractindextext

%
\IEEEpeerreviewmaketitle

\section{Introduction}

\IEEEPARstart{P}{ost-translational} modification (PTM) refers to the covalent modification of proteins following protein biosynthesis. It plays a vital role in a series of biological processes such as cell growth, proliferation, differentiation, metabolism and apoptosis, covering almost all life activities \cite{RN1649}. Since the last 20 years, the development of chemical probes and specific antibodies facilitates the enrichment for certain PTM, making the enrichment of modified peptides from cell proteomes possible. Moreover, the rapid development of high-resolution mass spectrometry (MS) has been providing an ideal platform for the identification of the PTM sites and quantification of their abundance and dynamic changes on a proteome-wide scale \cite{RN392}. Currently, more than 500 different kinds of PTMs have been identified \cite{RN964}, while cysteine (Cys) residues are often the most susceptible to modification.

\begin{figure*}[!hbt]
	\centering
	\includegraphics[width=0.8\linewidth]{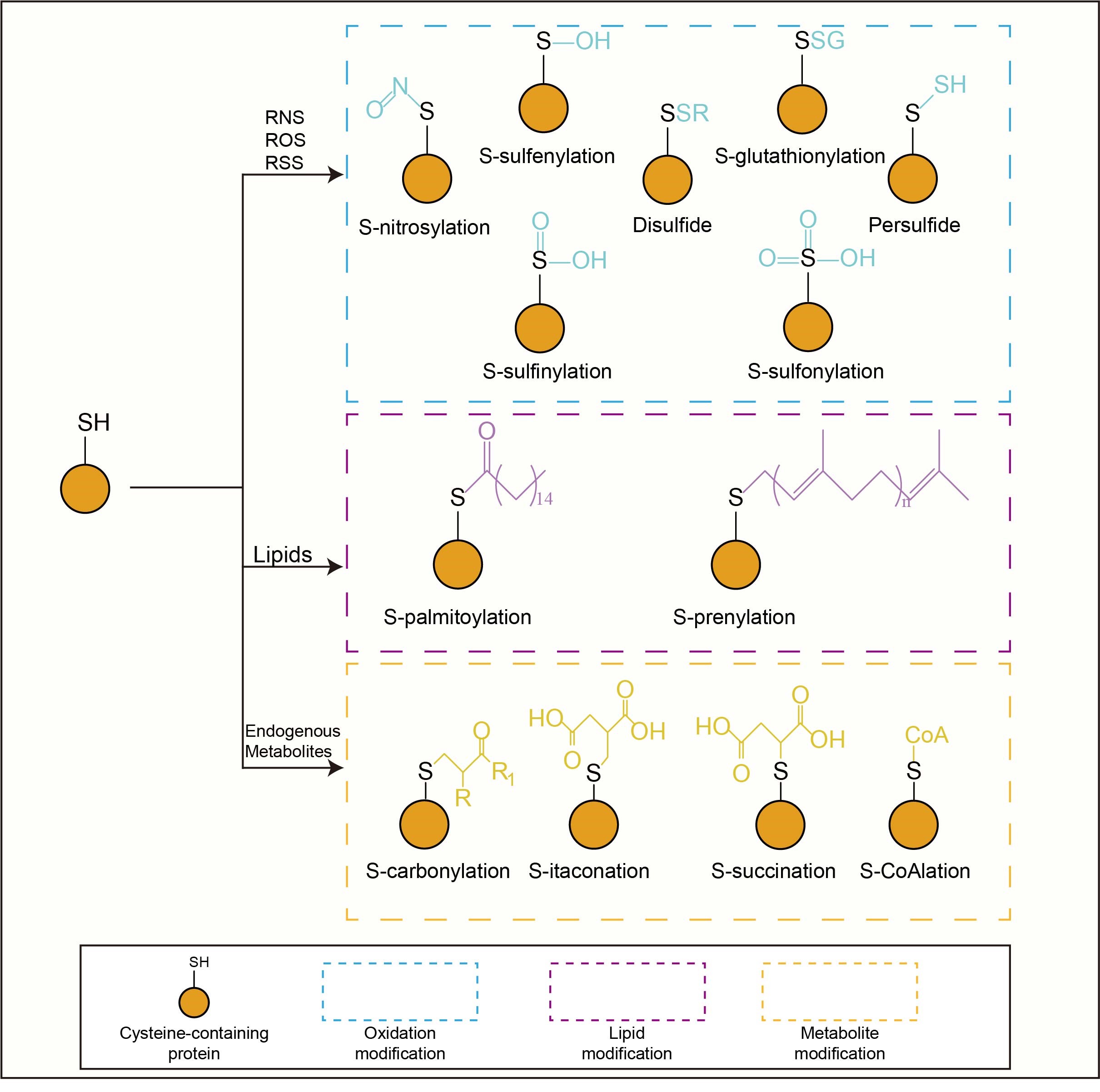}
	\caption{Modifications can be divided into three types according to their precursors.}
	\label{fig:f1}
\end{figure*}

In most proteome, cysteine residues occur low frequently but high in chemical reactivity. Due to the high reactivity, cysteine could be the nucleophilic residue attacking the substrate. Its thiol side chain could also react with each other and form a disulfide bond stabilizing the three-dimensional structures, changing the redox state, and chelate metal ions in some metalloprotein \cite{RN278}. Besides, different types of PTMs often occur in cysteine, which can be divided into three main types according to the difference of precursor molecules: oxidation PTM, lipid PTM and metabolite PTM (See Figure \ref{fig:f1}). These PTMs play important roles in regulating protein structures and functions and relate to a series of diseases \cite{RN1127}. Therefore, studies on cysteine PTM have important implications for both biology and health. In recent years, several aspects of biological MS have been improved especially measurement speed, sensitivity, resolution, and functional extension. Meanwhile, some new probes and enrichment methods were developed. With advances in these new technologies, more and more new PTM sites and related diseases have been identified and reported, like itaconation and succination \cite{RN1013}. These studies deeply revealed the regulation role for certain cysteine PTM in cell growth, proliferation, metabolism and differentiation. At the same time, as data science ramps up its applications in other fields, a range of prediction algorithms based on big data were used in bioinformatics, promoting the development of PTM prediction tools. These bioinformatical prediction tools combining proteomics data and machine learning expand our understanding of PTM and notably simplify the experimental implementation for modified cysteine site identification. Although some identification methods and related diseases for a single type of modifications have been reviewed, the panorama of whole cysteine PTM with research process at global proteomic level is still undescribed. In this review, we will focus on the recent advances in cysteine PTMs, covering three sections: 1) their enrichment processes, 2) proteomics research development and 3) prediction tools. We also describe the overlap between different cysteine PTM and provide future directions.

\section{Terminology and Background Concepts}

\begin{figure*}[!hbt]
	\centering
	\includegraphics[width=\linewidth]{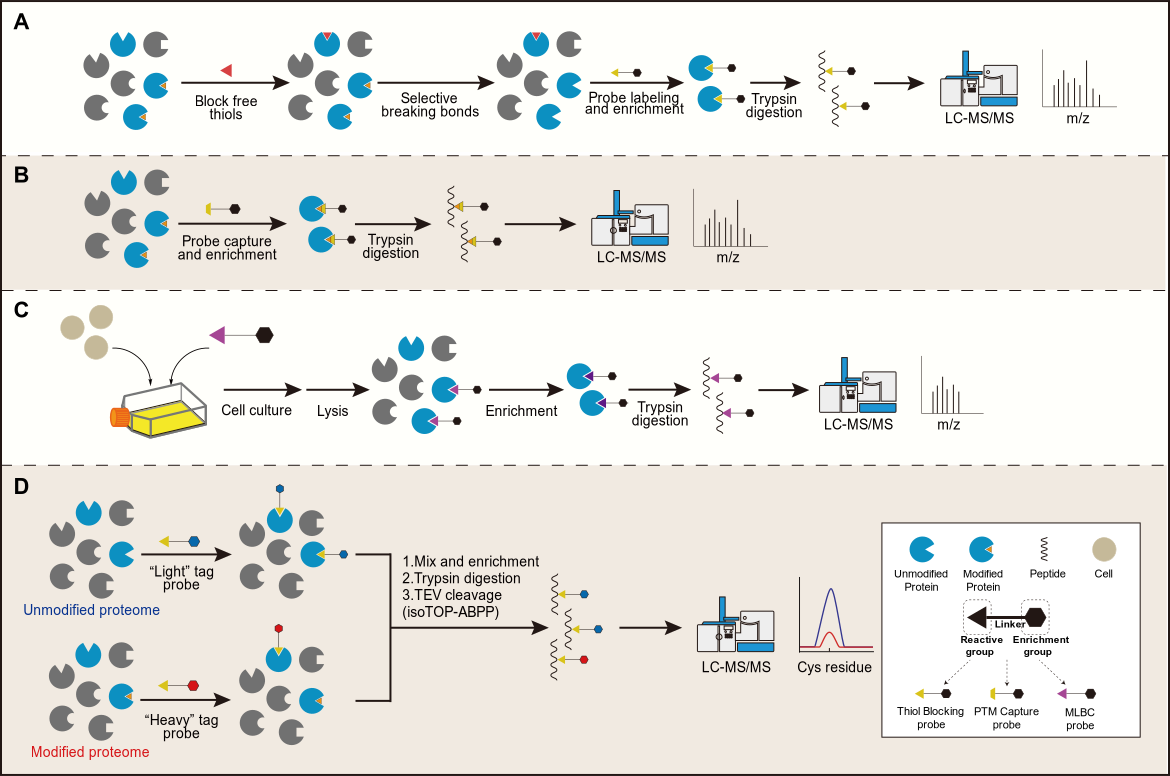}
	\caption{Brief processes of four enrichment categories for cysteine PTM enrichment: thiol blocking (A); Direct capture (B); Metabolic Labeling follow by Bioorthogonal Chemistry (C) and Thiol Isotope Labeling (D). These categories generally require probes for enrichment that include a reactive group linked to an enrichment group. The reactive group is bound to thiols or the cysteine PTMs and the enrichment group is employed for affinity purification.}
	\label{fig:f2}
\end{figure*}

Cysteine PTMs, like any other PTMs, are difficult to directly detect on the proteome-wide scale due to their highly dynamic nature and low abundance. Generally, antibodies for the enrichment of the modified proteins or peptides are useful but difficult to be developed. Alternatively, different strategies have been developed to enrich specific types of cysteine PTMs. We clustered them into four groups, according to different workflows: Thiol Blocking (Figure \ref{fig:f2}A), Direct Capture (Figure \ref{fig:f2}B), Metabolic Labeling follow by Bioorthogonal Chemistry (Figure \ref{fig:f2}C) and Thiol Isotope Labeling (Figure \ref{fig:f2}D). These strategies were detailed by a few reviews \cite{RN683, RN568, RN613}. In the following, we described and compared the strategies. 

\subsection{Strategy A: Thiol blocking}

The rationale of the thiol blocking strategy is the “blocking” of free cysteine thiols followed by conversion of cysteine PTMs via specific reagents to thiols, which are captured through reactive probes linked to enrichment groups (Figure \ref{fig:f2}A). The blocking reagents are generally N-ethylmaleimide (NEM) and iodacetamide (IAM) while the common enrichment groups include biotin \cite{RN276}. A series of reagents have been developed to break the PTM covalent bonds for oxidation and lipidation. For instance, ascorbate was used for the removal of S-nitrosylation (SNO), arsenite for S-sulfenylation (SOH), glutaredoxin for S-glutathionylation (SSG) and, DTT for all reversible oxidation PTMs whereas hydroxylamine could remove S-palmitoylation, respectively \cite{RN683, RN1114}. The main challenge of this strategy is the development of effective reagents that convert specific modifications to thiols. For example, such reagents for some metabolite PTMs (e.g. S-succination and S-carbonylation) have not been developed and these PTMs could not be enriched using this strategy.

\subsection{Strategy B: Direct capture}

In the direct capture strategy, tiny chemical probes are widely used to directly grasp specific cysteine PTMs (Figure \ref{fig:f2}B). The key of the probes is the reactive groups specific to PTMs. For instance, dimedone could detect the oxidation S-sulfenylation (SOH) \cite{RN891}, and hydrazide or hydroxylamine could capture the metabolite PTM S-carbonylation \cite{RN568}. The challenge of this strategy is the identification of reactive groups that react with cysteine PTMs with high specificity and sensitivity.

\subsection{Strategy C: Metabolic Labeling follow by Bioorthogonal Chemistry method}
Due to the lack of suitable probes for some cysteine PTM types and the increasing need to examine the PTMs in a native environment, a strategy based on PTM precursor-like probes has been established, named Metabolic Labeling follow by Bioorthogonal Chemistry (MLBC) (Figure \ref{fig:f2}C). As for a particular PTM, the related probes incorporate the PTM precursor with a bioorthogonal enrichment group, which can metabolically label proteins during cell culture in the formation of nascent PTM. These series of probes are expected to function with good cell-permeability, low cytotoxicity and competitive kinetics. Biotin was used as the enrichment group but later replaced by azide or alkyne (as a handle conjugated to an enrichment group using click chemistry) due to its large size with spatial resistance and potential destruction by intracellular enzymes \cite{RN961}. This strategy is widely employed to enrich lipid PTMs because their linkage with an enrichment group has little effect on their physicochemical properties and can pass through the cell membrane \cite{RN1114}. Moreover, this strategy is suitable for some cysteine PTMs that are difficult to be converted to thiols or directly captured, such as S-itaconation. Nevertheless, it is inappropriate for the detection of the PTMs that fail to pass through the cell membrane, such as S-CoAlation.

\subsection{Strategy D: Thiol Isotope Labeling }

The strategies described above required specific reagents or probes to directly target the different types of PTM sites. By contrast, in the Thiol Isotope Labelling (TIL) strategy, different isotopic probes were designed to target and quantify free cysteines of the proteomes under two distinct conditions where the PTM levels of interest are diverse (Figure \ref{fig:f2}D). Two common reagents are used: ICAT \cite{RN1135} and isoTOP-ABPP \cite{RN303}. ICAT consists of three elements: a reactive group (e.g. iodoacetamide), a linker that can incorporate stable isotopes and an affinity tag (e.g. biotin). IsoTOP-ABPP is composed of a reactive group and a handle (e.g. alkyne), which is conjugated to an enrichment group coupling with isotopically distinguishable tags using click chemistry. Paired samples from cells were treated with the isotopic reagents, pooled, enriched and analyzed by LC-MS/MS. The relative intensity ratio of light/heavy isotopic pairs used as a quantitative readout of relative cysteine-labeling stoichiometry (Figure. \ref{fig:f2}D). The ratio value of ~1 indicates that the cysteine was unaffected, whereas the extreme ratio value indicates the cysteine with modification change. The accurate measure of the TIL strategy relies on the conditions under which the modifications are significantly changed. For instance, the mutation of fumarate hydratase (FH) leads to high levels of intracellular oncometabolite fumarate that formed S-succination and the comparison of proteomes from FH -/- and FH+/+ cells were used to define “FH-regulated” cysteine residues \cite{RN1910}. This strategy can be used to identify which cysteine residues are modified and estimate what PTM types occur on the cysteine residues

\section{Oxidation PTMs}

\begin{figure*}[!hbt]
	\centering
	\includegraphics[width=\linewidth]{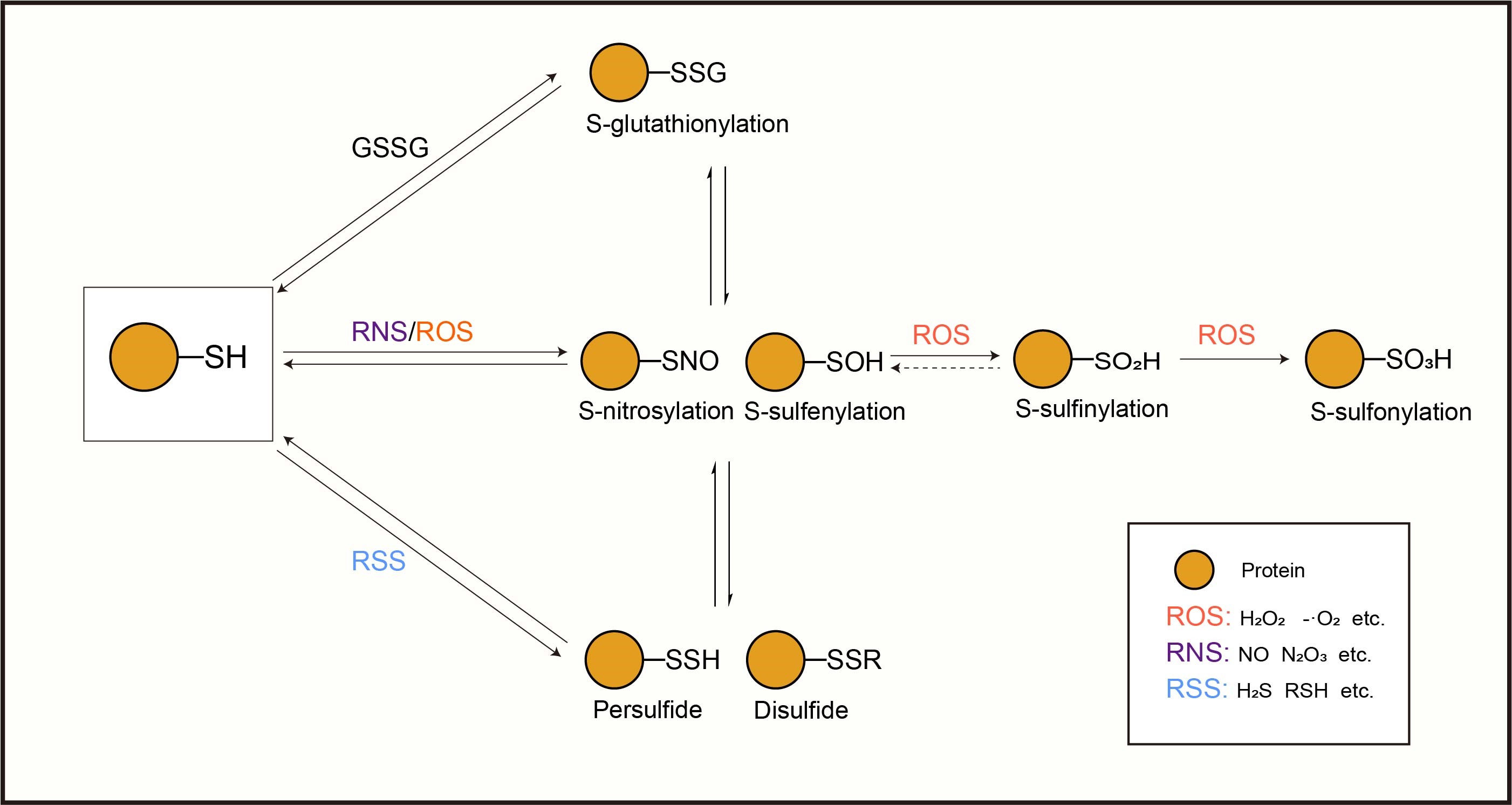}
	\caption{\label{fig:f3} The formation and conversion of different oxidation PTMs in proteome.Free thiols of cysteine residues react with RNS/ROS/RSS/GSSG and form revisable oxidation PTMs. They could result in irreversible S-sulfinylation and S-sulfonylation with the sustained ROS level. The switches between every type of reversible oxidation are under the regulation of some enzyme system.}
	
\end{figure*}

Redox reaction is an important type of chemical reaction in living organisms. Cysteine residues in proteins are easily oxidized by reactive oxygen species (ROS), reactive nitrogen species (RNS), reactive sulfur species (RSS) or GSH, due to high nucleophilicity of their thiols \cite{RN927} (Figure \ref{fig:f3}). Moreover, their oxidative sensibilities are affected by the flanking residues with different physical and chemical properties \cite{RN1127}. Generally, oxidation tends to occur on cysteine residues with lower pKa, higher solvent accessibility and shorter spatial distance to organelles with more oxidative species. The sulfur atom of cysteine residues in the -2 or +6 valence state is susceptible to oxidation. And these modifications are usually in in dynamic states under the regulation of some enzymes like Srx, Trx and Grx systems, switching to each other. Previously, this modification was thought as the product of oxidative injury. Recently it has been found to regulate protein functions via dynamically altering protein charges and steric hindrances \cite{RN893}. The regulation of protein functions via oxidation involves in many pathological events or diseases such as ageing, inflammation, oncogenesis and neurodegenerative diseases \cite{RN885, RN884, RN886}. Oxidations include both reversible and irreversible types according to whether it can be reduced by chemical reducing agents, which are described below, separately. 

\subsection{Reversible oxidation}

\subsubsection{Direct oxidation}

Cells could generate low amounts of ROS, RNS and RSS during normal aerobic metabolism, and increase the amounts under oxidative stress. Thiols with lower pKa tend to be oxidized by ROS or RNS, occurring reversible S-nitrosylation or S-sulfenylation, respectively. Both types of modifications could be further oxidized and become irreversible oxidation states like S-sulfinylation and S-sulfonylation. Thiols could also be converted into another reversible oxidation PTM named S-glutathionylation through the direct addition of oxidized glutathione (GSSG) \cite{RN686}. Additionally, some cysteine residues would be reversibly modified by RSS, forming disulfide or persulfide. These types of reversible oxidations are inter-transformable and such reversible switches regulate protein functions and affect signaling pathways.

S-nitrosylation (SNO) was first identified as cysteine oxidation in 1992 \cite{RN994}. It was considered as the product of the reaction between endogenous nitric oxide (NO) and cysteine-free thiol under the regulation of nitric oxide synthase (NOS) \cite{RN894}. It was later found to occur via other RNS such as N2O3 and GSNO. The association between SNO and biological processes or clinical diseases was summarized elsewhere \cite{RN1012, RN1011}. This modification is tough to identify without enrichment in proteome by LC-MS/MS due to lower abundance and dynamic change. A number of SNO sites were early detected on purified proteins or peptides through spectroscopic analysis methods based on the release of NO from SNO and MS-based methods with improved detection conditions \cite{RN999}. As this modification can be selectively reduced by ascorbic acid, a kind of Thiol Blocking method named biotin switch technique (BST) was developed for identifying S-nitrosylated proteins \cite{RN897}. The BST consists of a few principal steps: “blocking” of free cysteine thiols by some blocking reagents such as methylmethanethiosulfonate (MMT), formal reduction of SNOs to thiols with ascorbic acid, and in situ “labeling” the nascent thiols by a thiol specific biotinylating reagent like biotin-HDPP, followed by the purification by streptavidin for MS analysis. Based on this technique, a series of derivative methods were developed with advanced reaction systems and reagents. For instance, the BST-based method called SHIPS was developed that utilized a cysteinyl affinity resin to simultaneously “label” and "pull down" ascorbate reduced SNO-containing peptides \cite{RN898}. In this study, 162 SNO sites were identified from MDA-MB-231 cell lines, for which many of these modified proteins were enriched in cell death, protein degradation and mental ion binding pathways \cite{RN898}.

At the same time, a few chemical probes belonging to direct capture strategy were created to directly isolate and enrich the SNO peptides, such as gold nanoparticles (AuNPs) \cite{RN899}, mPEGb \cite{RN900} and TXPTS \cite{RN902}. For example, Doulias et al. used mPEGb to react with protein-containing SNO sites, finding more than 328 SNO sites belonging to 192 proteins. After the further analysis of modified proteins, they found these modified cysteine residues tend to locate in $\alpha$-helices and highly accessible surfaces bordered by other charged amino acids. 

In recent 5 years, the TIL strategy was gradually introduced in SNO research, which enabled the study of high accuracy and quantification. Zhou et al. studied SNO induced by GSNO-mediated transnitrosylation pathway in the MCF-7 cell line via isoTOP-ABPP \cite{RN903}. Among 610 identified SNO sites in this study, Cys58 in HADH2 and Cys329 in CTSD attracted the attention of researchers for their lack of annotation. Through the following experiments, these two cysteine residues were found that their nitrosylation can lead to reduced enzymatic activities despite they are not located in the enzyme activity center. Another important TIL method introduced to the SNO study is ICAT. Combining the Thiol Blocking method and ICAT, a new method called SNOxICAT was created by Chouchani and co-workers \cite{RN904}. They identified more than 1400 SNO sites in mouse cardiomyocytes, amounts of which localize to mitochondria. They also analyzed the relationship between SNO and the metabolism of cardiomyocytes under conditions of hypoxia and ischemia. Moreover, a series of quantitative methods combining the advantages of four chemical biology strategies have also been developed. Mnatsakanyan et al. created Cys-BOOST which enables high-precision quantitative research in the proteome \cite{RN878}. They got 8304 SNO sites in HeLa cells and presented a comparative sequence analysis to the flanking sequence of high-abundance SNO sites. The analysis results showed the enrichment of acidic amino acid residues in the flanking sequence. In accordance, they speculated the carboxyl groups on acidic amino acid residues can help the deprotonation of cysteine sulfhydryl groups and protonation of GSNO, which will promote SNO. 

Compared with nitrosylation, S-sulfenylation (SOH)is more complex. Cysteine residues tending to occur SOH often have extremely low pKa, some of them can even ionize to form nucleophilic thiolate (RS-) easy to react with peroxide to form SOH via bimolecular nucleophilic substitution. Besides, organic hydroperoxides (ROOH), peroxynitrous acid (ONOOH), peroxynitrite anion (ONOO-) and GSOH can also participate in the formation of SOH as hydroxyl group donor. \cite{RN683}. All these mechanisms make this modification more complex and dynamic, and it often becomes an intermediate for different oxidative modification transformations. 

The enrichment and identification technique for SOH is similar to the same strategy of SNO, however, due to the greater instability of SOH, a more direct and rapid identification method is particularly necessary. Similar to BST for studying SNO, SOH can be reduced by arsenite specifically and it can be used for SOH enrichment, but because of the cytotoxicity of arsenic salt and the high instability of SOH, the arsenite-based Thiol Blocking method was not widely developed for SOH research \cite{RN906}.

Dimedone and its derivatives can directly react with SOH, so the Direct Capture method based on this reaction is more widely used in the field of identification and analysis of SOH. A series of dimedone-based probes suitable for different reaction conditions have been developed \cite{RN907}. Carroll and co-workers used azide and alkyne instead of biotin to create a new series of probes named DAz and DYn, achieving better labelling results in living cells. They identified hundreds of SOH proteins that are sensitive to hydrogen peroxide in mammalian tumor cells and Arabidopsis cells, revealing the overall impact of SOH on cell life activities \cite{RN911, RN910}. They also introduced isotopes into DAz probes to achieve a quantitative comparison of SOH in normal and diseased cells\cite{RN912}. Yang et al. combined the tandem orthogonal proteolysis and DYn-2 probe, identifying more than 1000 SOH sites in human cell lines and found many sites localize in kinases, acetyltransferases, ubiquitin transferases and other PTM-related enzymes, which initially revealed crosstalk relationship of different modifications \cite{RN915}. In recent years, new identification techniques are still under development \cite{RN918}.

\subsubsection{Oxidation with disulfide bonds}

In addition to being oxidized by ROS and RNS, different thiol-containing molecules can also react with each other to form disulfide bonds (SSR). Apart from the earlier discovery of intra- and intermolecular disulfide bonds in proteins, the important regulatory roles of S-persulfation (SSH) and S-glutathionylation (SSG) in life activities have been gradually revealed in recent years. While SSH mainly plays a signaling role that directly alters protein activity, SSG modulates other redox modifications and participates in the protection against protein overoxidation \cite{RN919}. The functional roles of these PTM and their formation process have also been described \cite{RN935}.Some disulfide bonds can be formed within and between molecules while translating, playing a key role in maintaining the stability of protein structure and function under normal physiological conditions. Besides, other dynamic disulfide bonds can also regulate the formation and activation of multimeric proteins. The formation of a disulfide bond depends mainly on the spatial accessibility of the two sulfhydryl groups on cysteine. In early studies, two-dimensional electrophoresis was a classical research technique, but the low throughput of this method limits its efficacy on global proteome research. Thiol Blocking methods allow quantitative and accurate identifications in cells \cite{RN920}. He and co-workers created an efficient high-throughput MS technique named pLink-ss in 2015, which successfully identified hundreds of disulfide bonds in the secretory proteome of E. coli and human endothelial cells, achieving large-scale identification of disulfide bonds in complex biological systems \cite{RN921}.

Glutathione (GSH) is a small molecular tripeptide produced inside the cell, which has a reactive sulfhydryl group. This active sulfhydryl group enables them to bind to and scavenge intracellular oxidants and free radicals, maintaining a normal physiological environment in the cell. It can also bind to other thiol-containing proteins by forming intramolecular disulfide bonds or bind to another GSH molecule to form GSSG. The ratio of intracellular GSH to GSSG concentration is often in a dynamic equilibrium, and this mixed system can function as an oxidative buffer pair \cite{RN928}. GSH can also replace several other reversible oxidations and form S-glutathionylation (SSG), protecting cells when oxidative stress states occur. All of these functions make GSH involved in almost all biological redox reactions and many diseases associated with oxidation PTM. The first research method of this modification at the omics level on a large scale was MLBC. It coupled substances such as GSH, GSSG and GEE to biotin and studied the level of SSG after hydrogen peroxide stimulation of cells. Zaffagnini et al. identified 225 glutathionylated proteins in green algae through GSSG-based MLBC probe \cite{RN924}, while Chardonnet et al. used a similar strategy to identify more than 125 SSG sites in cyanobacteria for the first time and found them to be mostly related to oxidative stress regulation, carbon and nitrogen metabolism, and cell division \cite{RN923}. By expressing E.coli glutathionylspermidine synthetase (GspS) in the mammalian cell line 293T cells, Chiang et al. enabled cell to produce Gspm-biotin by catalyzing endogenous GSH to bind biotin-spm, which can be an intracellular probe identifying 1409 modification sites from 913 different proteins at once \cite{RN925}. This method circumvents the interference to endogenous modification molecules from the exogenous probes. Samarasinghe et al. took one step further by constructing a glutathione synthetase (GS) mutant to incorporate azido-Ala into glutathione, which can be used for enrichment by click chemistry \cite{RN929}. In recent years, MLBC has also been progressively expanded to large-scale proteomic assays in more model species and other mammalian cell lines \cite{RN930, RN932}. Since disulfide bonds formed by GSH can be specifically reduced by glutathione oxygen-reducing protein (GRX), GRX-based Thiol Blocking method has also been more frequently reported. Su et al. used GRX to convert SSG to thiol after alkylation closure and hydrogen peroxide stimulation of cell, then used thiol affinity resin combining with iTRAQ labelling to enrich and perform trypsin digestion, resulting in the identification of 1071 SSG sites in the mouse macrophage cell line RAW 264.7 cells \cite{RN926}. The study also quantitatively compared the extent of SSG modification after treatment with different concentrations of hydrogen peroxide and screened 364 highly sensitive cysteine sites, for which these proteins mostly associated with free radical scavenging and apoptosis. Li et al. combined this GRX-based Thiol Blocking method with iodoTMT to achieve proteome determination in Streptococcus mutants and identified 357 SSG sites \cite{RN931}.

In addition, Kramer et al. used another TMT label to quantify over 2000 sites in mouse skeletal muscle and found that more than half of these modification sites would change with exercise fatigue \cite{RN933}.
S-persulfation (SSH) was observed in the last century but has received less attention until the last decade. It can be obtained either by direct reaction of free sulfhydryl groups with sulfur-containing compounds or by transforming from other reversible oxidation PTMs. When SSH occurs, the newly added sulfhydryl group has a lower pKa, which makes it more reactive, and sometimes occur successive modifications called polysulfides \cite{RN935}. Due to the stronger reactivity of newly added thiols, it allows the preferential reaction of reactive substances with this new adding sulfhydryl group in a state of cellular oxidative stress, thus protecting the modified cysteine and preventing the cysteine from occurring irreversible oxidation. The main method to study this modification is Thiol Blocking method. In the early researches, the blocking agent MMTS and the breaking bonds agent DTT, which were not specific enough, had been used, causing some errors \cite{RN937, RN912}. Since SSH itself provides free sulfhydryl groups and the disulfide bonds in SSH are easily reduced, the new technology combined with the Thiol Blocking and Direct Capture has become the mainstream of research on SSH in the last five years. In 2016, Longen et al. developed qPerS-SID by improving the probe molecule and combined it with SILAC to achieve better quantitative detection of 782 SSH sites in mammalian cells \cite{RN940}. They also found a greater tendency towards aggregation for acidic amino acid residues around these sites through bioinformatic analysis and found more overlap sites with crosstalk by comparing them with several other cysteine oxidation. Gao et al. targeted ATF4, an important gene affecting sulfur dioxide (H2S) production, and identified more than 800 SSH sites closely related to ATF4 in MIN6 cells by controlling ATF4 biological activity and applying isotope-labelled NEM as a probe, revealing the relationship between SSH and the metabolic reprogramming of oxidative stress \cite{RN939}. Aroca et al. block all sulfhydryl groups by MSBT followed by CN-biotin-specific reduction, which is labelled to SSH sites one time. Then they analyzed SSH sites by mass spectrometry after biotin enrichment \cite{RN942}. The team used this strategy to identify 3174 different modified proteins in Arabidopsis, which were found to be mostly located in cytosol and chloroplast through gene oncology (GO) analysis and found to have more overlapping sites with SNO modifications, corroborating the crosstalk and interactions of both different oxidation PTMs. Wu et al. combined strong cation exchange (SCX) and ICAT to develop a new strategy called SSQPer, which identified hundreds of SSH sites in A549 cells \cite{RN941}. Fu et al. combined Thiol Blocking, Direct Capture and AIL method to develop QTRP, a direct quantitative assay \cite{RN938}. They applied it to detect 1547 SSH sites in a variety of mammalian cell lines and proved that some RSSH can be transformed with RSSR. Furthermore, proteome-level SSH modification studies have been extended to prokaryotes such as Staphylococcus aureus in recent years \cite{RN943}.

\subsubsection{Integration studies}
Although there are many types of reversible oxidation PTMs, these modifications are often interconvertible and in dynamic change, which makes it necessary to study them in a general and systematic way. Most of the cysteine reversible oxidation can be reduced by two reductants, DTT and TCEP, and a series of methods based on these reductants have been developed to help the understanding of the overall reversible oxidation. In 2018, Reest et al. created SICyLIA, an MS-based quantitative proteomics strategy for studying the overall effects of ROS on cells, enabling quantitative identification on MS by sealing thiols in experimental and control groups with IAMs carrying different isotopes and using DTT to reduce modification sites that have been oxidized followed by NEM sealing \cite{RN914}. They identified nearly 10,000 reversible oxidation sites in each of the acute and chronic oxidation models and found that chronic oxidation triggers metabolic adaptations in cells that balance the negative effects of oxidative stress on cells by increasing reductive substances. Topf et al. used the OxICAT technique to identify more than 4300 oxidation sites from yeast using TECP as a reducing agent in an H2O2-induced ROS environment and found that proteins with zinc finger structures and proteins involved in cellular transcriptional or translation processes are more susceptible to oxidation \cite{RN913}. Su et al. focused on adipocytes and identified more than 13,000 cysteine reversible oxidation sites in different tissues and cells by labelling with iodoTMT under the circumstance of drug-mimicking ROS \cite{RN1695}. They also chose insulin-treated cells to explore the crosstalk between oxidation and phosphorylation, revealing how upstream oxidized kinases of the phosphorylation regulatory network influenced phosphorylation levels throughout the cell. Xiao et al. designed a novel oxidative probe CPT, which combines TMT labelling with IMAC enrichment to achieve the largest reversible oxidative modification proteomics study to date. This team collected the identified sites by CPT and created Oximouse, a quantitative database of protein cysteine oxidation in various tissues and organs of mice including more than 34,000 non-repetitive modification sites. By performing analysis, they found that the elevated oxidized proteins in ageing mice are not directly caused by ROS, because most oxidized proteins are tissue-specific without the result of increasing ROS, thus overturning the previous understanding of the relationship between ageing and oxidation.

\subsection{Irreversible oxidation} 
When SOH is not reduced by intracellular regulatory systems in time or intense ROS and RNS persist in the intracellular environment, these cysteine sites can undergo irreversible S-sulfinylation (SO2H) and S-sulfonylation (SO3H) modifications. These two types of modifications cannot be reduced by conventional reducing agents in vitro and often lead to impaired function and protein degradation in vivo, so these modifications often seem like the end products of oxidation. However, some families of enzymes in some organisms, such as the sulfiredoxin (Srx) family, can reduce SO2H on some proteins to a reversible oxidation state by consuming ATP \cite{RN953}.
Enriching and studying these modifications is difficult due to the lack of reducing agents to break the chemical bonds. At the early stage of the research, there were some studies using chromatography to enrich these two modifications from the proteome. Lee et al. used the long column ultra-high-pressure liquid chromatography (UPLC)-MS/MS analysis and identified 61 irreversibly oxidized peptides \cite{RN948}. Paulech et al. used strong cation exchange resins combined with hydrophilic interaction liquid chromatography to purify and enrich irreversible oxidation from mouse cardiomyocyte proteome and identified 181 peptides occurring irreversibly oxidation \cite{RN949}. Subsequently, several probes that can react directly with these two modifications have been developed. Akter et al. developed a clickable electrophilic diazo probe based on dimedone backbone and named it DiaAlk, identifying a total of 387 irreversible oxidation cysteine sites from 296 proteins in A549 and Hela cells. By analyzing these identified sites, they also found that many of these modified proteins are located in exosomes and that the SO2H sites have more overlaps with the SOH sites \cite{RN947}.

\section{Lipid PTMs}
\label{sec:methods}

Lipid molecules are a large class of hydrophobic molecules widely presenting in all living organisms, which are often involved in the regulation of intracellular life activities, affecting many pathways including energy conversion, signal transduction, cell division, differentiation and so on. Lipid PTM or lipidation of proteins refers to the covalent bonds of lipid molecules to proteins, which could change their structure and function. Due to the nucleophilic sulfhydryl groups of cysteine, it can often form thioester and thioether bonds between cysteine thiol and some fatty acids or terpenes, respectively, resulting in lipidation of cysteine, of which the most common ones are S-palmitoylation (S-Palm) and S- prenylation (S-Pren).

\subsection{Palmitoylation}

\begin{figure*}[!hbt]
	\centering
	\includegraphics[width=0.8\linewidth]{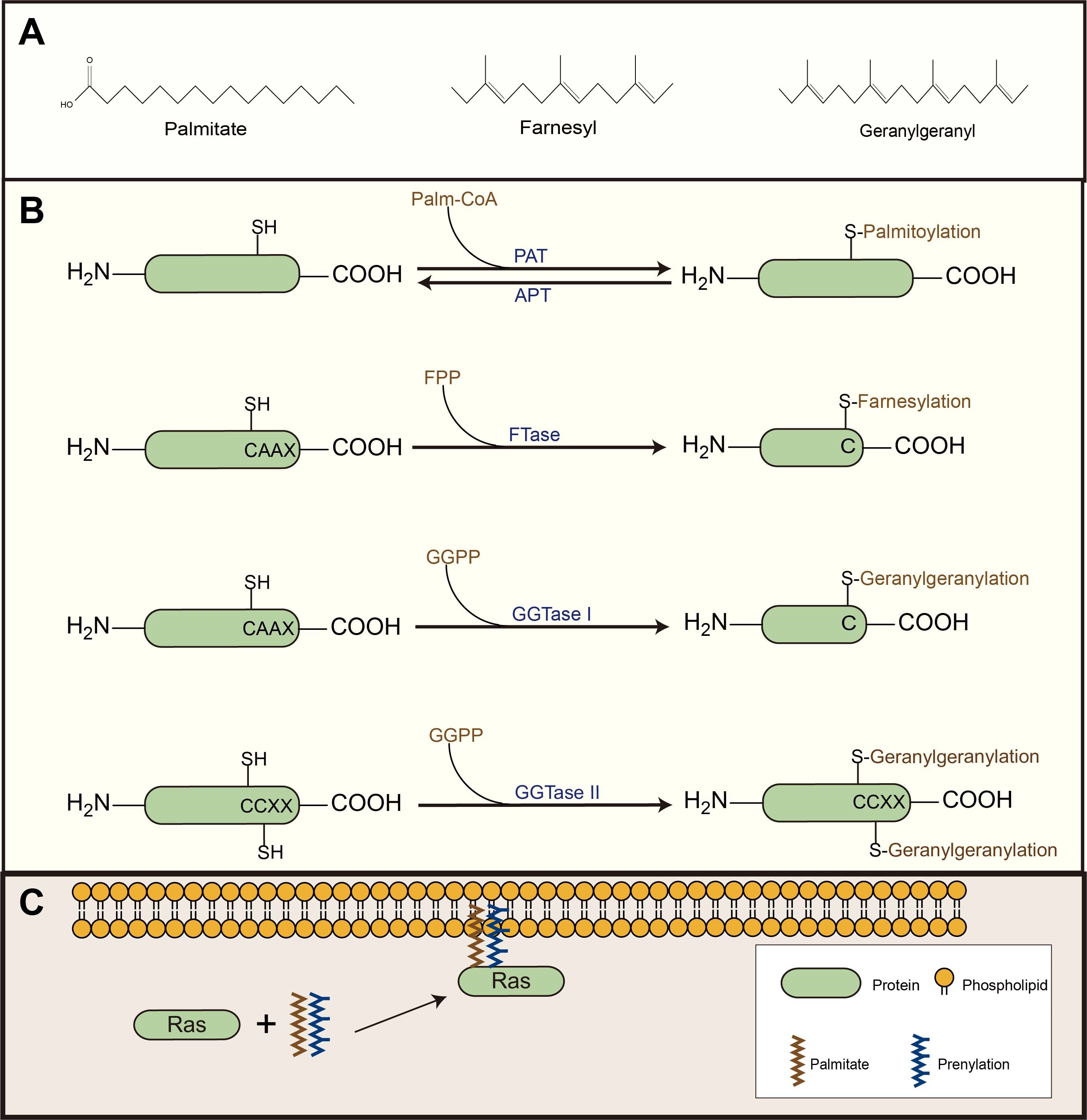}
	\caption{\label{fig:f4} Properties of lipids PTM.(A)Structures of three lipids. Farnesyl and Geranylgeranyl are both belong to isoprene.(B)Enzymatic reactions involved in cysteine lipids PTM.(C)Lipids PTM will increase protein hydrophilicity and makes them anchored to membranes easily, some proteins can exert their functions via this process such as Ras protein.}
	
\end{figure*}

A variety of fatty acids are naturally present in eukaryotic cells, some of which can covalently bond to proteins with some kinds of amino acids side chains, causing them to undergo acyl modifications. Palmitic acid, a 16-carbon saturated fatty acid, readily bonds to cysteines on proteins to form reversible S-palmitoylation (Figure \ref{fig:f4}A), a thioester bond. Protein acyltransferases (PATs) catalyze the formation of this modification, and most of these enzymes have a conserved Asp-His-His-Cys (DHHC) motif, which plays a key catalytic role, hence these enzymes are also named DHHC protein superfamily \cite{RN636}. And the clearance of S-palm is catalyzed by Acyl protein thioesterases (APTs) (Figure \ref{fig:f4} B). As a hydrophobic PTM, S-palm play a significant role in the   localization and stability of some membrane proteins such as MC1R and PD-L1, which are closely associated with tumorigenesis \cite{RN1137, RN1138}. Research on palmitoylation has been conducted since the last century, but due to the long-standing lack of commercial antibodies, such studies have been carried out mainly with different probes and chemical biology techniques \cite{RN613}. After decades of development, there is now a database called Swisspalm dedicated to S-palm proteins and sites \cite{RN641}.

There were two probe strategies applied to identify S-palm from proteome at the early stage of the research, one of which is known as the acyl-biotin exchange (ABE) and belongs to the Thiol Blocking strategy. It uses NH2OH to break the thioester bond, then captures the newly formed free thiol by biotin-HPDP \cite{RN642}. The other strategy uses long-chain fatty acid metabolism analogues containing alkyne or azide group to make MLBC probe \cite{RN643}. Based on these two approaches, S-palm at the proteomic level has been identified for more than seven species other than humans \cite{RN647}, including several important model organisms: yeast \cite{RN644}, rat \cite{RN645}, mouse \cite{RN648}, arabidopsis \cite{RN649}, etc. And these early studies have been reviewed \cite{RN633}. However, some kinds of MLBC probes are not specific enough and may trigger non-cysteine modifications, while Thiol Blocking method is likely to misidentify to non-palmitoylated thioester bond modifications, which limits the precision of earlier studies. Moreover, the lack of a suitable quantitative approach and the relatively low resolution of early proteomic identifications resulted in only small amounts of palmitoylation protein had been identified and many of them cannot be pinpointed to specific cysteine sites. Nevertheless, a series of important life activities regulated by palmitoylation, such as interferon-induced transmembrane protein 3 (IFITM3) mediated antiviral effects and Toll-like receptors (TLRs) mediated inflammatory responses were also revealed in these initial proteomic studies \cite{RN653, RN652}.

Over the past five years, a range of quantitative proteomics strategies has been introduced into the field of palmitoylation research, while more efficient enrichment methods and higher resolution MS techniques have made large-scale studies with precise identification of modified cysteine sites possible. Morrison et al. used an ABE approach combined with isotopic labelling to identify 280 palmitoylated proteins with their relative extent in human T cells \cite{RN655}. Collins et al. developed a site-specific Acyl-Biotin Exchange (ssABE) approach focusing on specific S-palm sites of mouse forebrain tissue and successfully identified 906 S-palm sites on 641 proteins \cite{RN606}. Many of these identified modified proteins belong to G-protein coupled receptors (GPCR) and ion channels, and their modified sites are mostly located on CC motifs. The researchers then hypothesized that palmitoylation could increase the hydrophobicity of such transmembrane proteins, thus making them more tightly bound to the cell membrane and facilitating their function. Zhang et al. combined ABE with cysteine stable isotope labelling to develop Cysteine-SILAC technology, achieving higher precision quantitative proteomics identification \cite{RN604}. In this study, two cell lines, MHCC-97L and HCC-LM3, were cultured in media containing different isotopes to obtain different isotopic compositions of polypeptide chains, which could be used for quantitative identification. A total of 2985 S-palm sites on 1443 proteins were identified in two cell lines, and there are 151 proteins with hugely different modification abundance. This study not only achieved quantitative identification of S-palm but also broke the limitation of the conventional SILAC technique that could not quantify peptides lacking K/R amino acids at the C-terminus. Subsequently, this team further improved the enrichment technique of palmitoylated proteins by replacing the biotin-HDPD probe with Fe3O4/SiO2-SSPy microsphere that can also form disulfide bonds and be enriched by the magnetic field, greatly improving the enrichment efficiency of the modified proteins \cite{RN634}. Through this strategy, a total of 1304 proteins were identified in the mouse brain tissue proteome, of which 797 proteins have not been previously reported to undergo such modifications.

Besides, as a reversible modification regulated by enzymes, the overall S-palm level in cells is always under a dynamic change. In 2011, Martin et al. applied SILAC and 17-Octadecynoic acid (17-ODYA) probe to BW5147-derived mouse T-cell hybridoma cells and compared the changes in overall cellular proteomic palmitoylation at different times \cite{RN656}. The results showed that the S-palm levels of dozens of proteins were in a more dynamic state, including Ras family GTPases, G proteins, etc., which are essential for cell life activities. Won et al. then designed a pulsed bioorthogonal S-palm assay to detect palmitoylation changes at different time points, achieving temporal analysis of S-palmitoylation dynamics by adding the depalmitoylase inhibitor HDFP to increase alkynyl fatty acid label \cite{RN607}. The team found that a significant proportion of S-palm levels did not change dramatically over time, implying that palmitoylation remained relatively stable although its removal was regulated by the enzyme. This type of research is still ongoing.

\subsection{Prenylation}

Isoprene (terpenoids) is another small molecule commonly involved in the lipid modification of proteins. It can react with cysteine sulfhydryl side chains in proteins and bind to proteins as more stable thioether bonds thus forming irreversible S-prenylation. This modification can be catalyzed by a class of enzymes known as prenyltransferase, including farnesyltransferase (FTase) that catalyzes farnesylation, and geranylgeranyltransferase (GGTase) that catalyzes geranylgeranylation. FTase and GGTase-I can specifically recognize substrate proteins with CAAX common sequence at the C-terminus of protein, where C represents cysteine, A represents aliphatic amino acids except alanine, and X is the amino acid that determines whether the protein undergoes farnesylation or geranylgeranylation \cite{RN662}. After the formation of S-pren, the AAX sequence is often excised, while GGTase-II catalyzes the addition of two geranylgeranyl groups to two cysteine residues near the C-terminus of the substrate protein having CXC or CCXX motif without enzymatic cleavage. This class of modifications can increase the hydrophobicity of modified protein and is often closely associated with the proper function of many membrane proteins, such as the signaling output of Ras protein \cite{RN665}. S-pren has also been found to be associated with the progression of chronic obstructive pulmonary disease (COPD), insulin resistance, fatty liver, tumors, Alzheimer's disease and a host of other diseases \cite{RN666, RN663}.

Since two kinds of S-pren can form two neutral losses of 204Da and 272Da on MS, most of the early proteomics studies investigating S-pren were based on this property. For instance, Wotske et al. used a reversed-phase chromatography to separate modified peptides from non-modified peptides and applied MS to identification \cite{RN670}, Bhawal et al. treated the modified peptides with oxidants and then distinguished farnesylation and geranylgeranylation by characteristic mass loss \cite{RN671}, etc. However, due to the lack of a sufficiently suitable protein enrichment strategy, the number of sites identified in this period of study was extremely low.

With the improvement of chemical proteomics technologies, more probes for identification and enrichment have been developed, especially the development of MLBC, advancing the histology of S-pren. Nguyen et al. attached biotin to a geranylgeranyl group and create biotin-geranyl pyrophosphate (BGPP) as an isoprene donor, allowing the enrichment of the modified protein from proteome species by avidin \cite{RN672}. Using this technique, the authors identified 42 proteins belonging to Rab GTPases that occur S-pren in the COS-7 cell line. However, the slightly larger molecular weight of the biotin and the potential to interfere with normal cellular metabolism have limited the application of such probes to larger studies, and the development of click chemistry-based probes has greatly decreased the potential for interference. Chan et al. first applied the GG-azide method to S-pren studies on proteomic level \cite{RN674}, while DeGraw et al. introduced more sensitive alkyne-based probes into the study, identifying dozens of proteins \cite{RN675}. Charron et al. identified nearly one hundred proteins in macrophage cell lines, including the zinc-finger antiviral protein (ZAP), which plays a key role in antiviral infection and immune activation \cite{RN338}. Suazo et al. identified a group of prenylated proteins of Plasmodium falciparum, which were shown to play important roles in the infection of cells by Plasmodium \cite{RN676}. In recent years, more studies have focused on the dynamic level of S-pren. Palsuledesai et al. combined MLBC probes with fluorescent moieties to achieve cellular labelling and Imaging of this modification in a variety of mammalian cell lines and found that it was indeed mostly located on the surface of membranous organelles\cite{RN673}. Storck et al. synthesized the isopentenyl-liked probe YnF and YnGG and combined the probes with a multifunctional capture reagent as well as quantitative chemical proteomic strategy. They investigated the suppressor effector of different classes of prenyltransferase inhibitors \cite{RN677}. They found that a class of proteins such as KRAS proteins could be modified by geranylgeranyltransferase when farnesyltransferase activity was inhibited, and eventually identified 10 Rab proteins with decreased levels of prenylation in a model of the retinal degenerative disease choroideremia.

Past studies had suggested that FTase preferred CAAX motif with X residues of alanine, serine, methionine or glutamine, while leucine, isoleucine and phenylalanine were preferred for GGTase-I. But in recent years, as the number of identified modification sites has risen, new evidence suggests that this pattern is not absolute and that many of the modified sites have a motif of C(x)3X \cite{RN668}. Related studies are still ongoing.

\section{Metabolite PTMs}

\begin{figure*}[!hbt]
	\centering
	\includegraphics[width=0.8\linewidth]{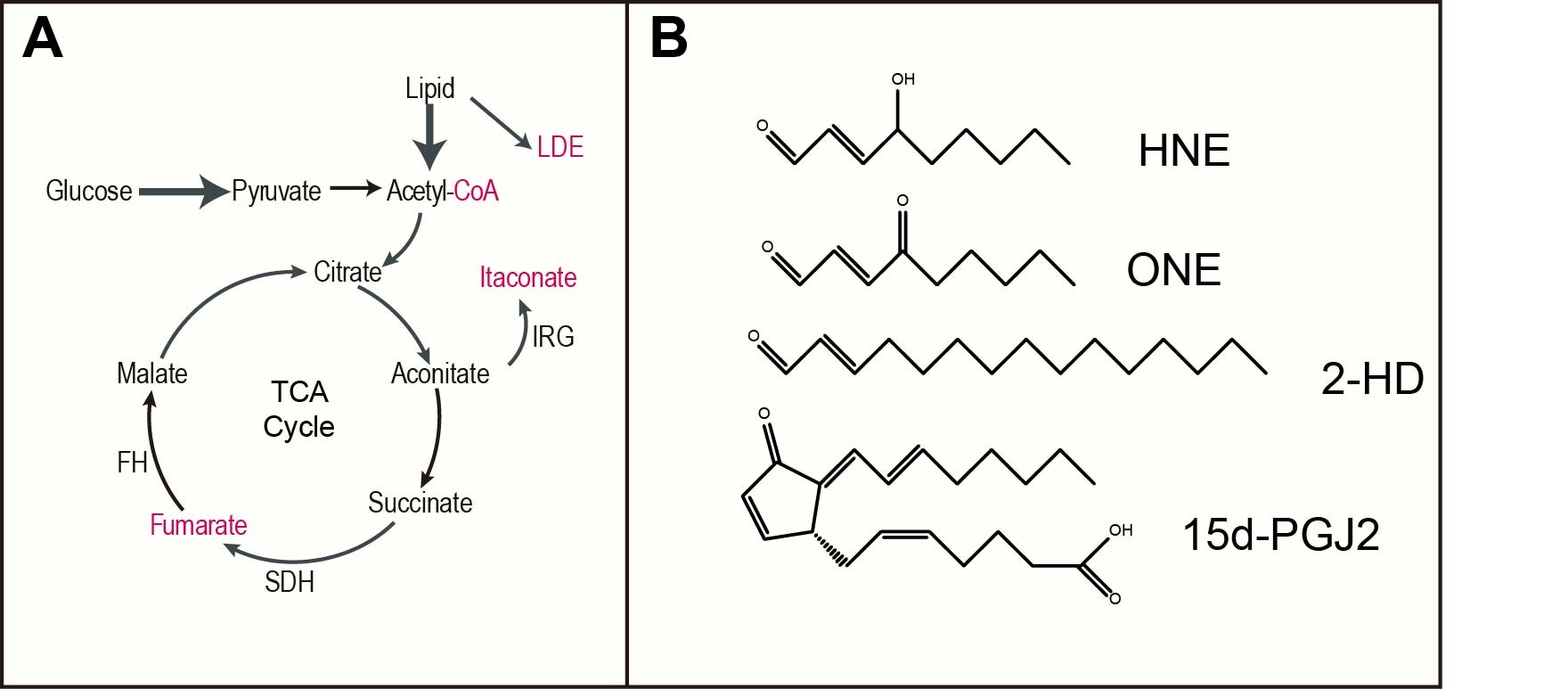}
	
	\caption{Overview of metabolic molecule.(A)The generations of different kinds of metabolic molecules in the metabolism with the enzyme regulation.(B)Structures of common lipid-derived electrophiles (LDEs).}
	\label{fig:f5}
	
\end{figure*}

Metabolism, as the most fundamental feature of living life, can directly meet the material needs for organism growth and development through generating energy and biomolecules. On the other hand, more and more studies have shown that some reactive metabolites such as Lipid-derived electrophiles (LDEs), fumarate, itaconate, coenzyme A and others, can also regulate cell signaling pathways through cysteine PTM (Figure \ref{fig:f5}A). Such modifications could occur at the low levels under physiological conditions, but trigger proteome-wide modifications under abnormal pathological or cellular stress conditions. These modifications are often non-enzymatic and irreversible, reducing the activity of many intracellular proteins thereby contributing to changes in signaling pathways \cite{RN1017}.

\subsection{Carbonylation}

When cells are under oxidative stress, the produced ROS often oxidizes many unsaturated lipids in the cell, producing Lipid-derived electrophiles (LDEs) such as 15-Deoxy-$\delta$12,14-prostaglandin J2 (15d-PGJ2), 2-trans-hexadecenal (2-HD), 4-hydroxy-2-nonenal (HNE), etc (Figure \ref{fig:f5}A). These LDEs tend to react with DNA or proteins without enzymatic mediation, resulting in irreversible carbonylation modifications \cite{RN569}. Cysteine is more susceptible to modification by electrophilic LDEs due to its nucleophilic sulfhydryl side chain, and such PTM may lead to changes in protein structure and functional activity. Although humans have discovered protein carbonylation for a long period and linked it to physiological processes such as stress, ageing, and apoptosis, the global proteomics studies did not begin until the last decade.

The main strategies to study this modification are Direct Capture methods, MLBC methods and TIL methods, while the applications of these methods in detail at the beginning of S-carbonylation proteome research have been reviewed \cite{RN568}.
In 2009, Codreanu et al. used Biotinamidohexanoic acid hydrazide to react with HNE-modified proteins, then followed by streptavidin capture, successfully identifying over 1500 HNE-modified proteins in RKO cells \cite{RN580}. This is the first HNE-modified human proteomic modification profile, but due to the low resolution of MS during the same time, this study was unable to locate detailed residue sites. Subsequently, a series of similar hydrazide-based Direct Capture probe continued to identify more S-carbonylation sites in different species and different cell lines \cite{RN586, RN587}, while some new probes based on other chemical groups that can readily react with unsaturated aldehydes were developed and applied to the study of S-carbonylation \cite{RN588, RN589}. During the same period, the first large-scale proteomic study based on MLBC was also made by Codreanu and co-workers \cite{RN602}. They treated two different cell lines, RKO and THP-1, with different concentrations of HNE and ONE analog probe to simulate the cellular damage caused by excess LDE. In total, more than 3000 modified proteins were localized and many of them were enriched in some important processes including protein synthesis, protein turnover and cytoskeleton regulation. However, the MLBC method still did not identify specific S-carbonylation sites until 2015, the introduction of the photo-cleavage group and isotopic labelling made it possible to probe specific S-carbonylation sites in the proteome \cite{RN603}. In that study, a total of 386 S-carbonylation sites were identified. Interestingly, the investigators found that some S-carbonylation sites can be more readily removed, suggesting there may be some unknown damage repair or regulation mechanism in cells. More precise site-specific studies are based on the TIL strategy. Wang et al. pinpointed over 1000 S-carbonylation sites in human proteome via isoTOP-ABPP and discovered different cysteine sites have distinct sensitivities to LDE. The study also selected a ZAK kinase newly identified in this experiment for cytological validation, demonstrating the modification on C22 of ZAK by HNE would suppress JnK activation in cells \cite{RN304}.

Most of the subsequent studies paid more attention to the S-carbonylation in different disease cell lines. Wang and co-workers developed a novel aniline probe based on MLBC with higher sensitivity than hydrazide or hydroxylamine probes, this probe is suited for studying endogenous S-carbonylation with low abundance \cite{RN567, RN571}. They firstly identified more than 1000 sites in non-small cell lung cancer (NSCLC) H1299 cells and 400 proteins in ferroptosis cell, building a solid foundation for subsequent in-depth mechanism exploration for NSCLC and ferroptosis. Some other subsequent studies focused on modifications triggered by a specific type of LDE. Xu et al. studied 2-HD-induced modifications and explored the functions of 2-HD via MLBC probe, identifying over 500 protein sites and experimentally verifying that the pro-apoptotic protein Bax can be covalently modified by 2-HD directly on the conserved Cys62 residue \cite{RN570}. Lu and co-workers developed a quantitative strategy with high sensitivity called six-plex isobaric labelling affinity purification (SiLAP), they identified and quantified 626 HNE-modified cysteine sites in complex proteomes \cite{RN572}. This series of proteomics studies have greatly expanded human knowledge about the overall life activity of cells related to LDE and provided new ideas for the treatment of diseases.

\subsection{Succination}

Fumarate is an intermediate product of the tricarboxylic acid (TCA) cycle. Sometimes it is also used in the food industry as a versatile food additive \cite{RN434}. In most cells, fumarate hydratase (FH) regulates fumarate content (Figure \ref{fig:f5}A), so its mutation will cause the excess accumulation of fumarate inside the cell and alter some cellular activities or induce a series of pathological changes. For example, it was shown that the accumulation of fumarate in cells can alter the epigenetic mechanisms and cytoplasmic pH \cite{RN629}. Moreover, fumarate can modify cysteine residues through Michael's addition reaction in the biological process named S-succination and this kind of PTM was first discovered in albumin and plasma proteins \cite{RN617}. With further research, a growing number of evidences have indicated the relationship between succination and many cellular activities, which suggests succination could represent good biomarkers for diseases including cancer and diabetes \cite{RN623}. Although an array of S-succination proteins were detected in both physiological and pathological conditions, people still poorly understood the overall S-succination in FH-deficient cells and their effects.

Several proteomics studies have driven marked advances in our understanding of S-succination. Naigai et al. used 2D gel electrophoresis, resulting in 60 S-succination protein fractions from the proteome of 3T3-L1 fibroblasts \cite{RN625}. They also identified 13 proteins by MALDI-TOF/TOF, firstly reporting this kind of PTM by MS-based method. Ming and co-workers undertook a proteomic-based screen in FH-deficient mouse embryonic fibroblast (MEF) cell line and renal cysts, identifying 94 protein S-succination sites containing Aconitase2 (ACO2) \cite{RN631}. Heterologous expression of human ACO2 in Fh1KO MEFs confirmed that three cysteine residues of ACO2 can be modified and cause an impaired enzyme activity upon high fumarate concentration, which could lead to abnormal intermediate content of TCA cycle and metabolic reprogramming. In 2019, Kulkarni et al. applied the isoTOP-ABPP-based AIL method to comprehensively and quantitatively analyzed cysteine reactivity changes in a human HLRCC cell line (UOK262, FH-/-), mapping 1170 cysteines sites occurred S-succination that most of them had not been detected before \cite{RN1910}. The modified proteins are mostly enriched in glycolysis, hypoxia, and reactive-oxygen stress pathways. After analyzing the flanking sequence of the modification sites, they found acidic amino acid residues are more tended to exist in the flanking sequence, which may have contributed to the increasing nucleophilic reactivity of the cysteine side chain through hydrogen ions dissociation. They also proved that excessive accumulation of fumarate caused by FH mutation could interfere with the formation of the SWI-SNF complex via modifying Cys520 of SMARCC1, and the complex is a significant tumor-suppressor.

In addition to cysteine residues occurring succination in FH mutant cells, a series of S-succination sites have been identified in other cell lines. Proli et al. identified several proteins modified by fumarate including tubulin,VDAC and DJ-1 in the brainstem cells of Ndufs4 KO mouse, a disease model of Leigh syndrome \cite{RN628}. Furthermore, it was found that accumulation of fumarate would induce mitochondrial dysfunction especially in the late stage of this disease. All of these phenomena may be associated with disease progression in individuals with this kind of neurodegenerative disease. Adam et al. identified more than 300 S-succination sites existing in pancreatic $\beta$ cells from diabetic mice lacking Fh1 and examined the relationship between S-succination and diabetes \cite{RN621}. It was unexpected that most of the mice lacking Fh1 did not present symptoms until 9 weeks,indicating that the metabolic bypass pathway may compensate for mitochondrial dysfunction caused by excessive fumarate. However, with the accumulation of fumarate in cells, the mitochondria would show severe morphological damage owing to a series of proteins are modified by fumarate such as GAPDH, GMPR, PARK7 and so on. A follow-up study confirmed that fumarate accumulation would promote the progression of diabetes as a consequence of the mutation of FH \cite{RN449}. Manuel et al. found that protein disulfide isomerase (PDI) in white adipose tissue of the diabetes model mouse would be modified by fumarate and showed a reduced activity \cite{RN616}. As a result, the secretion of pro-apoptotic protein C/EBP homologous protein (CHOP) and proinflammatory factors increased significantly, which then initiated endoplasmic reticulum stress. This research team also found the change of CHOP protein turnover was related to fumarate, revealing a new metabolic connection between anti-inflammatory adipocyte signaling and mitochondrial metabolic stress \cite{RN622}. Zhang et al. found fumarate could even modify glutathione and then induced sustained oxidative stress and senescence of the cell \cite{RN632}. Overall, these studies have explored the changes in metabolic pathways caused by S-succination, revealed the connections between these modifications and some diseases, providing novel therapeutic strategies to some diseases.

\subsection{Itaconation}

Itaconic acid is an $\alpha$,$\beta$-unsaturated carboxylic acid that was first discovered during the artificial distillation of citric acid. It has been widely used in industry, especially in the production of latex because of its ease of polymerization. However, it was not discovered to be an important product of metabolism and participated in the reprogramming of immune cells in mammals until 2011 \cite{RN453}. In macrophages, itaconate is produced from cis-aconitate, which is an intermediate product of the tricarboxylic acid (TCA) cycle (Figure \ref{fig:f5}A). This process is catalyzed by aconitate decarboxylase 1 (ACOD1) that was originally called immune-responsive gene 1 protein (IRG1) \cite{RN454}. With further studies, itaconate was confirmed to have anti-inflammatory functions via involvement in the negative regulation of immune effects especially suppressing inflammatory responses of macrophages and may have some pro-tumor effects \cite{RN464}. Itaconate and its derivatives are highly electrophilic, which makes it more likely to react with sulfhydryl groups of cysteine and form S-itaconation. However, under normal conditions GSH can neutralize the excess electrophilic material in the cell, thus, extensive S-itaconation at the proteomic level in cells can only be triggered by sufficiently high levels of itaconic acid, which is often induced by conditions associated with immune responses such as infections, tumors and stress responses \cite{RN455}. Although the study of itaconic acid in immunology has just begun in recent years, a series of studies on itaconic acid modification targets have helped researchers understand the important regulatory role it plays in the immune response \cite{RN985}.

In 2018, Luke and co-workers discovered that itaconate can act as a signaling molecule to block the inflammatory response of macrophages after many years of exploration \cite{RN518}. More than a dozen proteins modified by itaconate were identified by MS in this study, including Keap1, then they verified that the itaconation of Keap1 Cys151 would prevent the interaction between Keap1 and Nrf2, in the end, they also identified a negative regulatory loop between itaconate and IFN-$\beta$. In the following year, Kong and co-workers found GAPDH could also be modified by itaconate in M1 macrophages, which reduces the activity of GAPDH and blocks glycolysis, exerting an anti-inflammatory effect \cite{RN459}. Although both two studies have provided preliminary insights into the molecular mechanisms underlying the anti-inflammatory effects of S-itaconation, it is still not clear what global modifications are induced in the proteome of cells.

In 2019, Wang and co-workers developed a new Direct Capture probe, 1-OH-Az, which is based on glycosylation reactions, and combined it with competitive ABPP technology to identify 260 S-itaconation sites in the macrophage proteome for the first time on a large scale \cite{RN460}. They also found S-itaconation can significantly inhibit the glycolytic pathway of macrophages through covalent modification of Cys73 and Cys339 on ALDOA, thus decrease its activity and influence glycolysis. Therefore, regulating cellular energy metabolism through S-itaconation was considered as a new pathway mediating the anti-inflammatory effect. In the next year, the team developed a new MLBC probe called ITalk \cite{RN458}. Combined with this probe and quantitative chemical proteomics, 1164 S-itaconation sites belonging to 862 proteins were identified, some of which have never been reported before. They also focused on a new identified protein PIPK3, proving that the modification of its Cys360 will promote activation of its downstream signaling pathway and lead to cell necrosis. These proteomic studies provide rich data to understand the role of itaconate in the inflammatory response.

Notably, a new study has found that different derivatives of itaconate not only have different electrophilic reactivity but also have different target proteins, suggesting that subsequent studies need to pay attention to differentiating the biological effects of different derivatives or MLBC probe \cite{RN461}.

\subsection{CoAlation}

Coenzyme A is a class of small molecules widely existing in eukaryotic cells, and it is involved in almost all energy metabolic pathways in living organisms as a key cofactor for dozens of metabolic enzymes. Similar to GSH, Coenzyme A has a reactive thiol, but this thiol often binds acetyl groups to form acetyl coenzyme A, a very important acetyl group transmitter that participates in a variety of biosynthetic and biomodification related to acyl groups. Due to the presence of this sulfhydryl group, it can also modify cysteines in proteins, which is known as S-CoAlation firstly discovered in 1996 \cite{RN957}. However, due to the lack of appropriate enrichment and identification techniques, this modification has few corresponding studies.

With the successful development of specific antibodies in the last few years, it has become possible to study this modification at the proteomic level through the Direct Capture method. Tsuchiya et al. found a total of 123 peptides belonging to 91 proteins modified by coenzyme A in mouse heart and liver cells, and the increase in this PTM correlated with the level of cellular oxidative stress \cite{RN958}.  After analyzing the modified proteins, they found that more than half of them belong to metabolism-related enzymes, and the modification of CK, GAPDH, IDH and PDK2 affects their activities, thus demonstrating the regulatory role of this modification on cellular metabolism. Several years later, the team continued to dig deeper and identified 117 kinases that undergo this modification resulting in altered kinase activity \cite{RN1126}. They focused on the negative impact of S-CoAlation on Aurora A and its specific molecular mechanism. Currently, related proteomic studies are still in their infancy.

\section{Bioinformatics prediction}

\begin{figure*}[!hbt]
	\centering
	\includegraphics[width=0.8\linewidth]{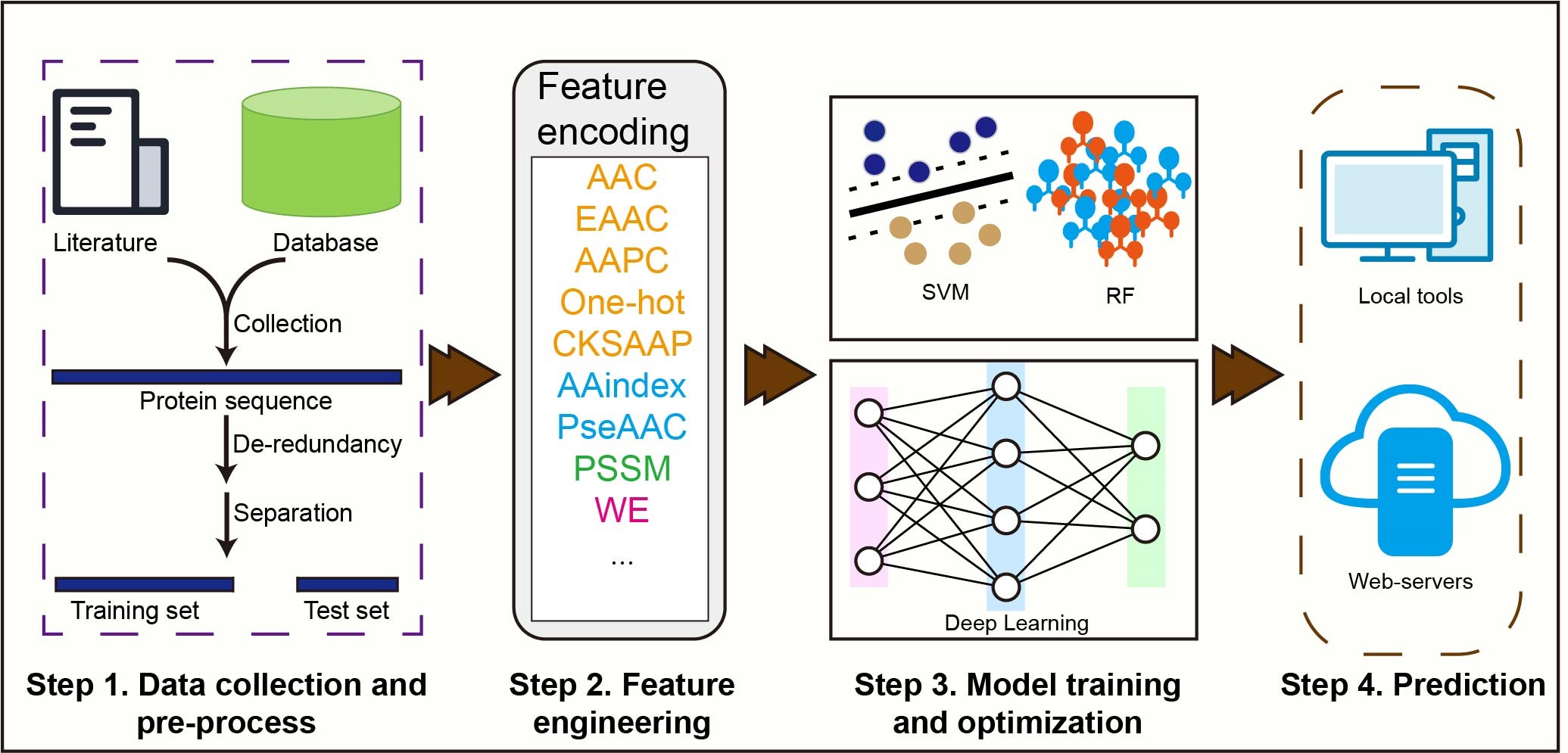}
	\caption{Brief flowchart template for bioinformatic prediction of cysteine PTM sites.The process of developing a prediction tool can be roughly categorized into four steps.}
	\label{fig:f6}
\end{figure*}

As the number of cysteine modification sites identified by chemical proteomics continues to increase, large quantities of related PTM data have been produced, which drives the bioinformatics study of PTM. These studies include not only functional analysis, but also predictive models using protein sequence as input. Researchers can develop bioinformatics tools to predict specific PTMs sites based on a range of algorithms and identified PTM sites. Such predictive models can be a complement to experimental approaches through screening for possible candidate modification sites and narrow the scope of experimental validation, especially when studying cysteine PTM-related signal transduction or protein function. In recent years, with the development of AI technology, prediction models based on machine learning algorithms with amino acid sequences feature input has become the dominant strategy in the field of cysteine PTM prediction. This strategy mostly consists of the following steps (Figure \ref{fig:f6}): (i) Data collection and pre-processing, (ii) Feature engineering, (iii) Model training and optimization, (iv) Prediction. Among all these steps, choosing appropriate types of feature coding methods and prediction algorithms are the most critical, which directly determine the final performance of the whole prediction model. Both of them should be chosen according to the data volume and data characteristics, and we will discuss the current common sequence feature coding method and prediction algorithms applied in cysteine PTM below.

\subsection{Feature encoding methods}
Machine learning algorithms require integer or floating-point number vector groups as input, and most of the cysteine PTM data formats today are character sequences, so they need to be encoded by sequence feature encoding techniques. An appropriate sequence feature coding method should optimize computational efforts while presenting the characteristics of the amino acids that make up the flanking sequence of PTM cysteine comprehensively and representatively. Dozens of sequence-based feature encoding approaches have been applied to the field of cysteine PTM prediction, which can be broadly classified into three categories: (i) Sequence Amino Acid Composition based (SeqAAC-based) encoding strategy, (ii) Amino Acid Physico-Chemical Properties based (AAPCB-based) strategy, (iii) Sequence Homology and Evolutionary Distance-based (SeqHED-based) strategy. We will focus on the most common methods used to summarize.

\subsubsection{SeqAAC-based strategy}
Amino acid composition (AAC) is the most classical type of sequence coding, which is based on the amino acid composition and arrangement of sequence. In this strategy, the frequency and location of amino acids at different positions around the modified cysteine are translated into numerical vector input that can be recognized by the training model. AAC is the most typical method of this strategy, which indicates the frequency of amino acid occurrences at each site of the sequence, and in recent years it has also been improved by sliding window and developed into the EAAC method \cite{RN1827}. Transformation of the sequences into matrices by representing amino acids as binary numbers is called one-hot or binary encoding (BE), which is another simple and effective method more suitable for vectorization operations \cite{RN1023}. Comparing amino acids at each pair of positions on the training set and constructing them into a matrix for input is an approach known as AAPC, which can compare both positive and negative sample data. The composition of the K-space amino acid pair (CKSAAP) combines the advantages of the above methods, it counts the frequency of each k-spaced (k = 0, 1, 2, 3, ...) amino acid pair in any given sequence. This method is very useful for finding modal motifs, such as the CXXC motif of highly reactive cysteines \cite{RN452}.

\subsubsection{AAPCB-based strategy}
Early encoding methods were mostly based on the SeqAAC-based strategy, but they could not better reflect the physical and chemical properties (PCB) of amino acids around the modified sites, such as hydrophobicity, molecular weight, polarity, pH, steric hindrance and other characteristics. Because the thiol reactivity of cysteine is affected by neighboring amino acids for their side chain, incorporating the PCB of amino acids into feature encoding can increase the amount of information and thus improve the accuracy of prediction. AAindex is the most commonly used AAPCB-based method, which is a database containing more than 500 indices based on quantified PCB of amino acids, each index contains 20 values representing 20 amino acids \cite{RN611}. Another feature encoding method is pseudo amino acid composition (PseAAC), which adds the weight value of PCB to AAC, accommodates both operations and information \cite{RN1026}. Both of these two methods have more derivative methods. Since the reactivity of cysteine residues is easily influenced by the surrounding environment, the AAPCB-based strategy tends to perform well for cysteine modifications.

\subsubsection{SeqHED-based strategy}
Some evolutionarily conserved flanking sequences of cysteine sites may contain specific motifs or amino acid residues that play a key role in promoting or maintaining the occurrence of PTM, which are shown by the high tendency of the same type of PTM occurring in these sequences. Several encoding methods based on sequence homology or evolutionary distance (SeqHED) can effectively improve the prediction accuracy of the model. The most commonly used SeqHED-based approach is the position-specific scoring matrix (PSSM), it reflects the probability that the amino acids around the PTM site would change to other amino acids during evolution \cite{RN1034}. Each residue in the amino acid fragment in this feature corresponds to 20 different amino acids getting different conserved states, which can form an L × 20 scoring matrix (L is the sequence length). Besides, some other methods K nearest neighbor algorithm (KNN) combining with BLOSUM62 were sometimes used in feature encoding. These methods also have many derivative methods.

\subsection{Prediction model algorithm}
\subsubsection{Traditional machine learning}

Machine learning (ML) is a widely-used data analysis and prediction technique in the field of data science, which contains a very large number of modelling algorithms. By converting the modified and unmodified sequences into training sets, the PTM prediction problem can be transformed into a classification problem in machine learning. Most of the ML algorithms used for the prediction of cysteine PTM in the last five years have focused on either Random Forest (RF) or Support Vector Machine (SVM), while SVM is more popular than RF in the cysteine PTM prediction field.
RF is based on the decision tree algorithm, which is a decision model with a tree structure based on the attributes of the data. RF randomizes the use of variables and data, thus generating a large number of random trees. For each input, the entire random tree is traversed, and the maximum number of judgments is taken as the final output \cite{RN1024}. SVM, on the other hand, is an approach specifically designed to solve classification problems. Its principle is to map points in low-dimensional space to high-dimensional space so that they become linearly divisible, and then determine the classification boundary via linear division. Although RF models are highly interpretable and can reflect the biological context through well-trained models, their overall performance is often inferior to that of SVM \cite{RN1028}.
Also, some other algorithms such as logistic regression, Bayesian discriminant algorithm, general artificial neural network (ANN), KNN combining with dipeptides and other methods have been applied to the study of cysteine PTM, but their principles are more complex or require extremely high tuning parameters, so they have not become mainstream approach, and we will not discuss them here.

\subsubsection{Deep learning}
Deep learning (DL) is an algorithm that emulates the neural activity of the human brain, developing from ANN. Due to its unique working principle, the manual feature engineering of traditional machine learning for sequence encoding can be simplified and replaced with word embedding (WE), automatically selecting the optimal features. But it requires more data for training. With the increasing amount of identified cysteine PTM sites, some prediction models have started to apply deep learning in recent years \cite{RN974}.
There are three types of deep learning commonly used for cysteine PTM prediction: Recurrent Neural Network (RNN), Long Short-Term Memory (LSTM), and Convolutional Neural Network (CNN). RNN is a class of neural networks that perform repetitive recurrent operations over time, which allows them to have some kind of "memory" of the information that has been recirculated, making them very suitable for classification problems of sequences carrying biological information. LSTM can be regarded as a special kind of RNN, which reinforces the memory of specific sequence information, and it will determine whether different information will be forgotten or remembered to continue passing on according to the characteristics of the constituent elements. CNN uses convolution instead of matrix multiplication, achieving feature extraction and recognition judgment of the input information through a large number of intermediate convolutional and pooling layers. It is more suitable for image processing, but can also handle biological sequences.

To improve prediction accuracy, many current studies also combine more than one encoding method and modelling approach to build composite models. Several integrated sequence extraction and modelling tools have also been developed to help other researchers not familiar with bioinformation to study cysteine PTM. We also summarize the prediction tools for various PTM of cysteine for the researchers in need (Table \ref{table:datasets}).

\begin{table*}[!t]
	\centering
	\caption{A comprehensive summary of reported prediction tools}
	\resizebox{\linewidth}{!}{
		\begin{tabular}{|c|c|c|c|c|c|}
			\hline
			\textbf{Tool}      & \textbf{PTM type}         & \textbf{Data set} & \textbf{Algorithm} & \textbf{Website}                                  & \textbf{Reference}   \\ \hline
			DeepCSO            & S-sulfenylation           & 1535/8819         & LSTM               & http://www.bioinfogo.org/DeepCSO                  & \cite{RN1827}       \\ \hline
			fastSulf-DNN       & S-sulfenylation           & 900/6856          & DNN                & https://github.com/khanhlee/fastSulf-DNN          & \cite{RN1020}       \\ \hline
			GPS-Palm           & S-palmitoylation          & 3098/18992        & CNN                & http:// gpspalm.biocuckoo.cn/download.php.        & \cite{RN1021}       \\ \hline
			DeepGSH            & S-glutathionylation       & 1688/1957         & DNN                & http://deepgsh.cancerbio.info/                    & \cite{RN1022}       \\ \hline
			SIMLIN             & S-sulfenylation           & 1235/9349         & RF, SVM            & http://simlin.erc.monash.edu                      & \cite{RN1023}       \\ \hline
			PreSNO             & S-nitrosylation           & 3383/17165        & RF, SVM            & http://kurata14.bio.kyutech.ac.jp/PreSNO/         & \cite{RN1024}       \\ \hline
			-                  & S-nitrosylation           & 731/810           & SVM                & -                                                 & \cite{RN1025}       \\ \hline
			SPalmitoylC-PseAAC & S-palmitoylation          & 436/500           & ANN                & www.biopred.org/palm                              & \cite{RN1026}       \\ \hline
			CapsNet            & S-palmitoylation          & 2589/572          & CapsNet            & https://github.com/duolinwang/CapsNet\_PTM        & \cite{RN1027}       \\ \hline
			Sulf\_FSVM         & S-sulfenylation           & 900/6856          & SVM                & 123.206.31.171/Sulf\_FSVM/                        & \cite{RN1028} \\ \hline
			SVM-SulfoSite      & S-sulfenylation           & 900/6857          & SVM                & https://github.com/HussamAlbarakati/SVM-Sulfosite & \cite{RN1029}       \\ \hline
			PredCSO            & S-sulfenylation           & 228/757           & GTB                & -                                                 & \cite{RN1030}       \\ \hline
			SulCysSite         & S-sulfenylation           & 1027/8023         & RF                 & http://kurata14.bio.kyutech.ac.jp/SulCysSite/     & \cite{RN1040}       \\ \hline
			MDD-Palm           & S-palmitoylation          & 710/5676          & SVM                & http://csb.cse.yzu.edu.tw/MDDPalm/                & \cite{RN1031}       \\ \hline
			iPreny-PseAAC      & C-terminal Prenylation    & 94/456            & SVM                & http://app.aporc.org/iPreny-PseAAC/               & \cite{RN1032}       \\ \hline
			S-SulfPred         & S-sulfenylation           & 900/6858          & SVM                & -                                                 & \cite{RN1033}       \\ \hline
			RSCP               & Redox-sensitive cysteines & 758/758           & SVM                & http://bio- computer.bio.cuhk.edu.hk/RSCP         & \cite{RN1034}       \\ \hline
			SOHPRED            & S-sulfenylation           & 1031/8028         & Bayesian,RF, SVM          & http://genomics.fzu.edu.cn/SOHPRED                & \cite{RN1035}       \\ \hline
			iSulf-Cys          & S-sulfenylation           & 900/6856          & SVM                & http://app.aporc.org/iSulf-Cys/                   & \cite{RN1041}       \\ \hline
			SOHSite            & S-sulfenylation           & 1145/8368         & SVM                & -                                                 & \cite{RN1036}       \\ \hline
			PGluS              & S-glutathionylation       & 2325/13234        & SVM                & http://59.73.198.144:8088/PGluS/                  & \cite{RN1037}       \\ \hline
			iSNO-ANBPB         & S-nitrosylation           & 1229 /1223        & SVM                & -                                                 & \cite{RN1038}       \\ \hline
		\end{tabular}
	}
	\label{table:datasets}
\end{table*}

\begin{figure*}[!hbt]
	\centering
	\includegraphics[width=0.8\linewidth]{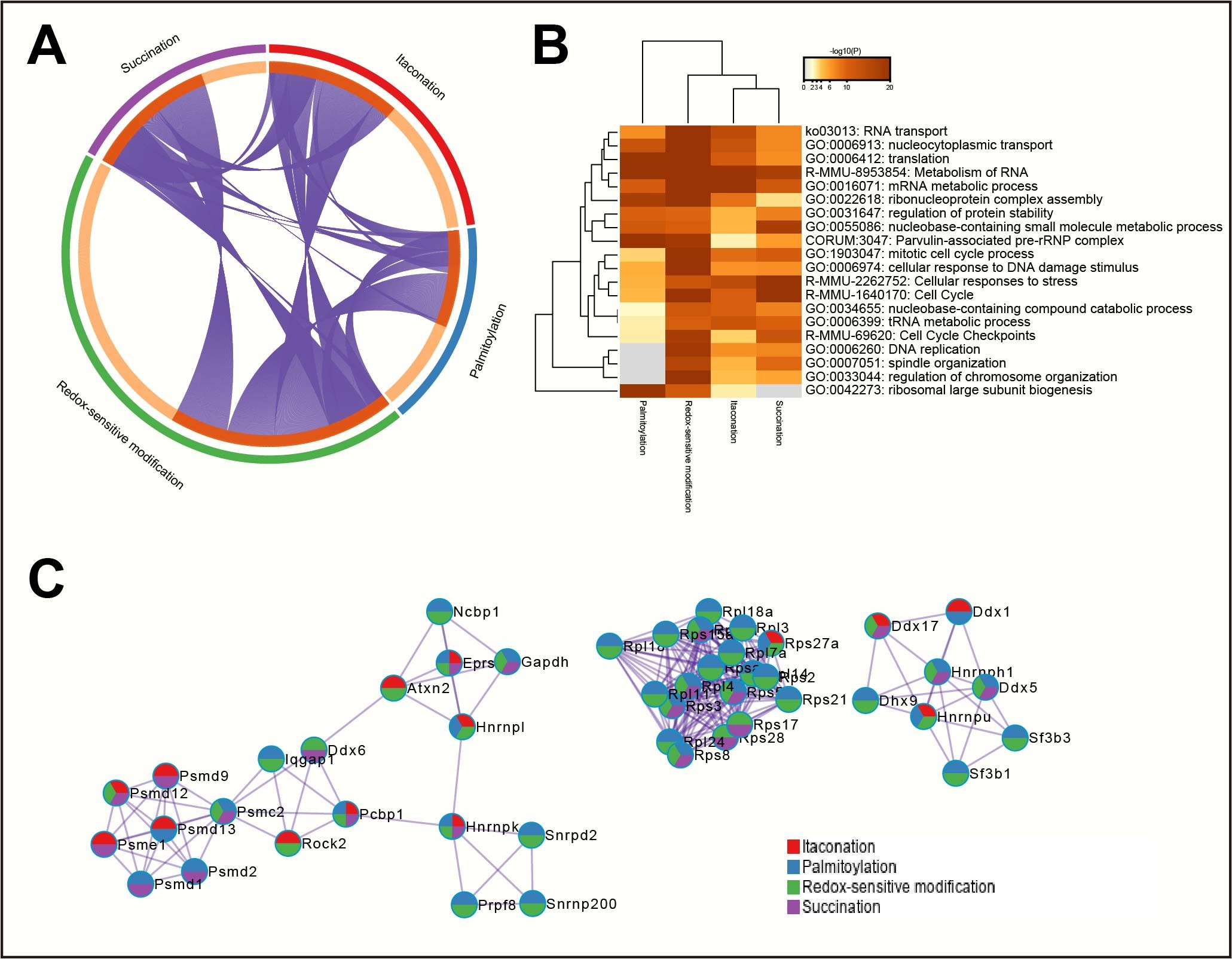}
	\caption{Modifications can be divided into three types according to their precursors.(A)Overlap of co-identified cysteine PTM proteins. (B)Heatmap showing the top enrichment clusters of cysteine PTM protein.(C)Top three best p-value MCODE components identified from the merged network, which are related to RNA Metabolism.}
	\label{fig:f7}
\end{figure*}

\section{Perspective, Summary, and Future Directions}

Cysteine can undergo such a wide variety of PTM because of its unique physicochemical properties, and these different PTMs can sometimes even occur simultaneously and in crosstalk with each other under specific conditions. For instance, if macrophages are activated under inflammation and undergo metabolic reprogramming, not only can intracellular ROS contribute to the occurrence of oxidation PTMs, S-Carbonylation and S-CoAlation (named together as Redox-sensitive modification), but their production of large amounts of itaconate can also cause S-itaconation in the proteome. Through Metascape, we have performed a systematic analysis of multiple cysteine PTM that have been identified in mammalian cells \cite{RN980}. The results show that many cysteine-containing proteins can undergo more than one modification, suggesting the possibility of crosstalk and even competition between these modifications (Figure \ref{fig:f7}A). Enrichment analysis of these protein functions, biological processes and responses show that the metabolic pathways or molecular functions are mainly enriched to RNA metabolism and cellular stress-related pathways (Figure \ref{fig:f7}B). These data suggest that cysteine PTM on the proteome may act as a regulatory network in synergy with mechanisms such as reprogramming of cellular metabolism and selective expression of genes under specific conditions, so it plays an important role in regulating life activities in specific physiological or pathological states of cells. Through PPI interaction network analysis, the three most enriched clusters are all associated with RNA metabolism, which may reveal an evolutionarily conserved mechanism beyond one kind of PTM that regulates cellular life activities. The past studies of individual cysteine PTM mostly pay much attention to cellular responses to the environment and energy metabolism, while the analysis of co-modification sites takes one step further to explore the crosstalk of these PTMs and their place in biological evolution (Figure \ref{fig:f7}C). Similarly, some studies have begun to link cysteine PTM to alterations of other modifications such as phosphorylation and ubiquitination at neighboring sites. In the future, probing the interplay network of different PTMs could help us understand how this rapid regulation work at the protein level without significant alterations in gene expression, and could contribute to the understanding of some of the abnormal PTMs caused by disease development and thus develop effective therapeutic interventions.

Cysteine has a highly reactive thiol and undergoes a variety of PTMs, but the abundance of these PTM is often low and in a dynamic flux. In the last decade, a series of chemical proteomics strategies have been developed that successfully expanded human understanding of these modifications at the proteomic level. At the same time, the large amount of data made it possible to build bioinformatics prediction tools using data science methods. Experimental validation after the prediction of target proteins changed the research process. Some important proteins modification, such as the palmitoylation of PD-L1, were thus identified with the help of a prediction tool \cite{RN639}.

In addition to abnormalities in some endogenous cysteine PTM that may be associated with cellular pathological states or diseases, some exogenous electrophilic small molecules similar to endogenous molecules can be covalently modified with cysteine, resulting in altered cellular states and leading to disease \cite{RN981}. Some drugs also modify many intracellular cysteine-containing proteins and affect the biological activity of some key proteins, thus exerting their drug effects \cite{RN307}. This type of research is still in its infancy and is a direction that researchers should work on in the next phase.

In the future, it still needs to develop new proteomic identification technologies and bioinformatics prediction tools to expand human knowledge of cysteine PTMs. Many identification techniques require adding probes to cell lysates, which may cause partial modification changes and do not enable the observation of dynamic changes in a given modification within cells under specific conditions, so it is necessary to develop highly sensitive intracellular labelling techniques. Some chemical agents, like thiol blockers and breaking agents that were once widely used in the past, may occur certain side reactions and would cause false positives or false negatives, disturbing the experimental results, which also places an increased demand on the specificity of next-generation labelling techniques. Many chemical proteomics assay strategies are too time-consuming and costly, so there is a general trend to develop shorter and more cost-effective assay strategies. Moreover, among all three kinds of cysteine PTMs, metabolite PTMs is still in its infancy, needing more attention. In the end, with the development of data science and the iteration of computing hardware, more powerful algorithms and arithmetic have led to the gradual applications of DL-based prediction methods for PTM prediction. Almost all existing prediction tools are focused on oxidation PTMs and use mostly traditional machine learning strategies. It is also important to develop more efficient and accurate prediction tools to cover various modifications of cysteine by making full use of existing data.

\bibliographystyle{IEEEtran}
\bibliography{references}

%




\end{document}